\documentclass{elsarticle}

\journal{Journal of Management Science and Engineering}

\usepackage{amsmath,amsthm}
\usepackage{amssymb}
\usepackage{graphicx,setspace}
\usepackage{amsmath,amssymb}
\usepackage{color}
\usepackage[english]{babel}
\usepackage{mathrsfs,dsfont}
\usepackage{array}
\usepackage{tikz}
\usepackage{adjustbox}
\usepackage{multirow}
\usepackage{float}
\usepackage{hyperref}
\usepackage{booktabs}
\usepackage{graphicx}
\usetikzlibrary{arrows, automata, decorations.pathmorphing, backgrounds, positioning, fit, petri, matrix, patterns}
\usetikzlibrary{shapes,snakes}

\usepackage{lineno}
\usepackage{rotating}
\usepackage{pdflscape}

\usepackage[colorinlistoftodos]{todonotes}

\usepackage{setspace}
\usepackage{mathrsfs}
\usepackage[mathscr]{euscript}
\usepackage{bbm}

\usepackage{mathtools}

\usepackage{subcaption}
\usepackage{inputenc}

\newtheorem{theorem}{\bf Theorem}

\newtheorem{definition}{Definition}

\def\QED{~\rule[-1pt]{5pt}{5pt}\par\medskip}

\let\ALP  \mathcal

\DeclareMathOperator{\argmax}{argmax\ }

\DeclareMathOperator{\maximize}{maximize}
\DeclareMathOperator{\minimize}{minimize}
\DeclareMathOperator{\subjectto}{s.t.\ }
\DeclarePairedDelimiter\floor{\lfloor}{\rfloor}

\usepackage[linesnumbered,vlined,ruled]{algorithm2e}

\makeatletter
\def\endthebibliography{%
    \def\@noitemerr{\@latex@warning{Empty `thebibliography' environment}}%
    \endlist
}
\makeatother

\begin{document}

\begin{frontmatter}

\title{Multi-Objective Vehicle Rebalancing for Ridehailing System using a Reinforcement Learning Approach}


\author[mymainaddress]{Yuntian Deng\corref{mycorrespondingauthor}}\ead{deng.556@osu.edu}

\author[mymainaddress]{Hao Chen}\ead{chen.6945@osu.edu}

\author[mymainaddress]{Shiping Shao}\ead{shao.367@osu.edu}

\author[mymainaddress]{Jiacheng Tang}\ead{tang.481@osu.edu}

\author[mymainaddress]{Jianzong Pi}\ead{pi.35@osu.edu}

\author[mymainaddress]{Abhishek Gupta}
\cortext[mycorrespondingauthor]{Corresponding author}
\ead{gupta.706@osu.edu}


\address[mymainaddress]{Department of Electrical and Computer Engineering, The Ohio State University, Columbus, OH 43210, USA}


\begin{abstract}
The problem of designing a rebalancing algorithm for a large-scale ridehailing system with asymmetric demand is considered here. We pose the rebalancing problem within a semi Markov decision problem (SMDP) framework with closed queues of vehicles serving stationary, but asymmetric demand, over a large city with multiple nodes (representing neighborhoods). We assume that the passengers queue up at every node until they are matched with a vehicle. The goal of the SMDP is to minimize a convex combination of the waiting time of the passengers and the total empty vehicle miles traveled. The resulting SMDP appears to be difficult to solve for closed-form expression for the rebalancing strategy. As a result, we use a deep reinforcement learning algorithm to determine the approximately optimal solution to the SMDP. The trained policy is compared with other well-known algorithms for rebalancing, which are designed to address other objectives (such as to minimize demand drop probability) for the ridehailing problem. 
\end{abstract}


\end{frontmatter}

\section{Introduction}

Mobility-on-Demand (MoD) systems, developed by transportation network companies (TNCs), are believed to make the urban transportation system more efficient and affordable.  Uber, one of the leading companies, has achieved over \$65 Billion in gross bookings in 2019, with a 31\% year-over-year growth \cite{website2019}. Such systems often induce an imbalance between the vehicle supply and the travel demand in some areas, especially during rush hours. Customers usually experience high prices and long waiting times while the drivers move around the city waiting for ride requests in low demand areas. This leads to fuel wastage and high passenger waiting times. Such issues increase the operational costs of the drivers and fleet managers and reduce the quality of service for the passengers. As a result, TNCs experience an instant loss in revenue because of the imbalance and a loss in business over time as fewer drivers and customers will enter the market if the imbalance persists for a long duration. 

Thus, rebalancing vehicles to address the imbalance is one of the major problems faced by ridehailing systems around the world. In this problem, assuming the number of drivers is sufficiently large, there is an inherent trade-off between customer waiting time and Empty Vehicle Miles Traveled (EVMT). Systems with small passenger waiting time usually will have significantly larger EVMT, thereby increasing the operational cost for the drivers (on average). Further, it is not feasible in practice to know in advance the exact time of customer arrivals, and thus it is difficult to make predictions and relocate vehicles in advance. To address the last issue, past demand data can be used to construct a probabilistic model for demand distribution over the city. This model can guide the choice of rebalancing policies to minimize the expected operational cost and at the same time, provide a reasonable expected waiting time for the customers.  


Accordingly, in this paper, we model the rebalancing problem as a semi Markov decision problem (SMDP) over a complete graph, in which nodes are the neighborhoods of the city and the edges represent the road network with the associated travel time. In our model, we assume that the passengers arrive at every node according to a Poisson distribution with a known arrival rate. Every 100 seconds, the TNC platform makes rebalancing decisions.
Since this rebalancing control is applied every 100 seconds, it is a semi Markov decision problem. The rebalancing policy of the platform takes as input the current imbalance in the supply and the demand, and determines the rebalancing control action to be taken under that state; as a result, this is a {\it state dependent} rebalancing policy. The cost function for the SMDP is taken to be a convex combination of the waiting time of the passengers and the EVMT. The goal of the platform is to determine the optimal rebalancing policy.

Since the problem is difficult to solve analytically, we use reinforcement learning to determine an approximately optimal state-dependent rebalancing policy. We developed two tools that helped us use the reinforcement learning algorithm:
\begin{enumerate}
    \item We built a data processing tool that takes as input the New York taxi trip data from the month of December 2019 (a total of 7 million trips) and uses statistical techniques to extract the arrival distribution of the passengers from the Manhattan area of New York. The details of this data processing tool are provided in \ref{app:dataprocess}.
    \item We built a simulator that takes as input the arrival distribution of the passengers at each node, the rebalancing policy, and the fleet size of the vehicles, and runs a Monte Carlo simulation to compute the waiting time of passengers and EVMT traveled under that rebalancing policy. The code for this simulator was written in Python.
\end{enumerate}
We use proximal policy optimization (PPO) with a deep neural network as a value function and policy function approximator to determine the approximately optimal state-dependent rebalancing policy. We also compare the performance of the trained rebalancing policy with other rebalancing policies in the literature and show through Monte-Carlo simulations that the rebalancing policy designed using PPO outperforms the other rebalancing policies. 



The rest of the paper is organized as follows. Section \ref{sec:review} reviews the current literature on rebalancing policies and reinforcement learning-based approaches for transportation systems. The SMDP problem is formulated Section \ref{sec:model}. Section \ref{sec:algorithm} discuss the PPO algorithm and the experiment setup is presented in Section \ref{sec:setup}. In Section \ref{sec:results}, we present the simulation results and compare the trained policy with the other rebalancing policies. Finally, Section \ref{sec:conclusion} concludes this paper and presents some directions for future work.

\section{Related Work}\label{sec:review}

We review some related literature in this section.

\subsection{Rebalancing Methods}

Devising rebalancing policies for TNCs has received significant attention over the past decade. We classify them into two parts based on methodologies: state-independent policy and state-dependent policy.

The state-independent policies balance the vehicle distribution based on the average dynamics of the system. Pavone et al. \cite{Pavone2012} utilized a fluid model for vehicles and develop a rebalancing policy to ensure that the system reaches equilibrium, at which there are empty vehicles at every node and no waiting customers at each node. The rebalancing policy is posed as a solution to a Linear Programming (LP) problem that minimizes the cost of rebalancing. Smith et al. \cite{Smith2013} proposed a mobility-on-demand system mixing both customer-driven vehicles and taxi service. With a requirement of bounded waiting customers, both the number of vehicles and the number of drivers are minimized by solving two minimum cost flow problems. Rossi et al. \cite{Zhang2016} also introduced a flow-based network model that is congestion sensitive. The rebalancing policy is devised by solving an LP that minimizes the total trip time and fleet size with the congestion entering the constraints of the LP. Sayarshad and Chow \cite{Sayarshad2017} proposed a new queuing-based formulation for relocating idle vehicles and develop a Lagrangian decomposition heuristic. Considering estimated future states, the proposed algorithm outperforms the myopic policy when minimizing the operational cost and queuing delay. The queuing network model is also used in Zhang et al. \cite{zhang2016control} and Iglesias et al. \cite{Iglesias2019}. The rebalancing process in \cite{zhang2016control} is treated as a virtual arrival process, which is independent with respect to the real arrival process. The rebalancing policy is an LP that minimizes the total rebalancing cost while keeping an identical demand-supply ratio (also known as utilization) among all the nodes. The rebalancing policy developed in \cite{Iglesias2019} also considers the congestion on the road as a constraint in the LP formulation of the rebalancing cost optimization problem.

However, the above policies are based on the steady-state analysis since they use either a fluid model or the queuing model. The steady-state analysis can only provide asymptotic optimality and fails to react to the current state and the dynamical evolution of the system. It also leads to a state-independent randomized rebalancing control policy. Moreover, the policies in \cite{zhang2016control,Iglesias2019} also require impatient passengers who quit the system if not matched to a vehicle immediately, since it simplifies the system dynamics and allows us to carry out the steady-state analysis.


State-dependent policies, where rebalancing control is dependent on the current queue lengths of empty vehicles and waiting passengers, can achieve significantly better performance. Banerjee et al. \cite{Banerjee2018} introduce Scaled MaxWeight rebalancing policy, which is a state-dependent rebalancing policy. This paper shows that no state-independent policy can achieve exponential decay of demand dropping probability with respect to the fleet size, while naive state-dependent policy (MaxWeight) achieves exponential decay without knowing demand arrival rates. Similarly, Kanoria and Qian \cite{Kanoria2019} propose Mirror Backpressure policy based on an increasing function, called score function, of queue length, allowing for more general demand arrival rates. These two policies assume that the vehicles can move from one node to another within one timestep, which is inappropriate in a city with large and asymmetric trip times. This modeling assumption is satisfied in a restricted downtown area, where the distance between two locations is generally small.   


\subsection{Reinforcement Learning}
The usual approach for solving MDPs is value iteration or policy iteration \cite{puterman2014markov}. However, if the state space is large or the value or policy functions cannot be stored in a compact form, then reinforcement learning (RL) has emerged as a remarkable tool for devising approximately optimal policies \cite{sutton2018reinforcement}. RL algorithms use data from a simulator or a real-world application to determine optimal decision policies in MDPs. Since in MoD systems, the state and action spaces are large and the value and policy functions difficult to store, RL with an appropriate function approximator is a suitable tool for devising approximately optimal rebalancing policies.

An et al. \cite{an2019rebalancing} model the rebalancing problem in the car-sharing system as a Markov Decision Process and set picking bonus and parking bonus for users as action. The state is defined as the number of remaining vehicles at each station. Deep Deterministic Policy Gradient algorithm is used to approximate the value function as the action space is continuous.

Wen, Zhao, and Jaillet \cite{wen2017rebalancing} utilize the deep Q network for rebalancing the shared mobility-on-demand systems. The state space is defined as the distribution of idle vehicles, in-service vehicles, and the predicted demand (passenger arrival rate). Each vehicle either stays put or moves to the highest value neighborhood (computed appropriately). One case study of London shows the deep Q network performs effectively by reducing the waiting time of customers. However, future demand is difficult to estimate in practice, and relying on this information makes it less realistic to be implemented.

Gueriau and Dusparic \cite{gueriau2018samod} propose a Q-learning-based decentralized approach to vehicle relocation as well as ride request assignment in shared mobility-on-demand systems. The full agent state consists of (i) vehicle state (empty, one passenger or full), (ii) presence of currently active requests in agent's zone (yes, no), and (iii) presence of currently active requests in neighboring zones. Three basic actions an agent can execute are defined as pickup, rebalance, and doNothing. Both ridehailing (one request by one vehicle) and ride-sharing (multiple requests by one vehicle) cases are evaluated through the demand generated from the NYC taxi dataset. However, the rebalancing strategies are limited in their setting, where agents can only choose one of four existing heuristic policies and may not execute the optimal rebalancing action.

Motivated by these studies, we design a more general rebalancing strategy, which requires only the passenger queue and vehicle queue length, without the need for demand prediction. Also, the control policy directly specifies the number of empty vehicles to be dispatched and the destination of these vehicles. This allows us to explore a wider class of policies, and as simulation suggests, gives rise to a better rebalancing strategy. We design our total cost as a convex combination of vehicle rebalancing miles and passenger waiting time with an adjustable parameter that controls the relative loss associated with each component. These two components represent the major trade-off in the system: the operational cost of EVMT v.s. quality of service. By combining these two, service providers can fine-tune the algorithm to fit their specific preferences.

\section{System Model}\label{sec:model}

The city is divided into $n$ regions. We model the city as a complete graph $\mathcal{G}=(\mathcal{N},\mathcal{E})$, where $\mathcal{N}=\{1,...,n\}$ is the set of nodes/stations and $\mathcal{E}=\{(i,j) | i, j \in \mathcal{N} \}$ is the road connecting the two stations $i,j$. Station $i \in \mathcal{N}$ is the centroid of the corresponding region. We let $t_{ij}$ denote the deterministic travel time between origin node $i$ and destination node $j$. We assume that $t_{ij}$ and $t_{ji}$ can take different values, depending on the traffic and road conditions in the city.  

Each passenger enters the system with a starting station $i \in \mathcal{N}$ and destination station $j \in \mathcal{N}$. Passenger arrival for each origin and destination pair $(i,j)$ is assumed to follow a Poisson process with the arrival rate denoted by $\lambda_{ij}$. We assume that $\lambda_{ii} = 0$. Such an assumption on the arrival process is cross-validated by the New York Taxi data, and the details are presented in Section \ref{sec:dataproc} and \ref{app:dataprocess}. The interarrival time between two passengers is assumed to be independent of all other randomnesses in the system. At each station $i$, passengers who originated at this station form a First-Come-First-Serve (FCFS) queue regardless of their destinations. 

The city is a closed system with respect to $m$ vehicles, that is, a fleet of $m$ vehicles are serving the demand. At every time step, each vehicle can have only two statuses: Either be traveling between two stations or wait at a station with no passengers. If the vehicle is traveling between the two stations, then it can transport a passenger or be rebalanced. Each vehicle can carry one passenger at most, and we assume that there is no ride-sharing between two distinct passengers. 

\subsection{System Dynamics}
We now introduce the notations to capture the system dynamics. Let $p_i[t]$ and $v_i[t]$ denote the number of passengers and vehicles at station $i$ at time step $t$ before the platform matches the passengers and vehicles. Let $p_{ij}[t]$ be the number of passengers who are waiting to reach destination $j$. Let $b_{ij}[t]$ denote the number of passengers arriving at station $i$ with destination $j$ during time interval $[t-1,t)$. Further, we have $p_i[t]=\sum_{j\neq i} p_{ij}[t]$ and  $b_i[t]=\sum_{j\neq i} b_{ij}[t]$. Let $x_{ij}[t]$ and $y_{ij}[t]$ denote, respectively, the numbers of non-empty vehicles and empty vehicles dispatched from $i$ to $j$ at time step $t$, respectively. The number of passengers leaving station $i$ is $x_i[t]=\sum_{j\neq i} x_{ij}[t]$.

It is further assumed that waiting passengers get matched as soon as there are empty vehicles at the same station, i.e., the number of passenger departing node $i$, $x_i[t]$ is the minimum of the current passenger queue length $p_i[t]$ and the current number of vehicles available $v_i[t]$ at that station. This leads to the following expression

\begin{align} \label{eqn_xit}
x_{i}[t] = \min \{p_i[t], v_i[t] \} \quad\text{ for all } i \in \mathcal{N}.
\end{align}

Since there are at most $v_i[t]$ vehicles available, the total number of vehicles leaving a station regardless of transporting passengers or rebalancing is upper bounded by

\begin{align} \label{eqn_xityit}
 x_{i}[t] + y_{i}[t] \leq v_i[t] \quad\text{ for all } i \in \mathcal{N}.
\end{align}

Combining \eqref{eqn_xit} and \eqref{eqn_xityit}, the constraint on the number of empty vehicles dispatched from each station is

\begin{align}
    y_i[t]\triangleq\sum_{j\neq i}y_{ij}[t]\leq\max\{v_i[t]-p_i[t],0\}
\end{align}

The total number of vehicles arriving at station $i$ within the time interval $[t-1,t)$ is given by $\sum_{j\neq i} (x_{ji}[t-t_{ji}]+y_{ji}[t-t_{ji}])$. For each station $i$, the number of waiting passengers $p_i[t]$ can be updated recursively based on passenger departures $x_i[t-1]$ and arrivals $b_i[t]$ within one time slot. Similarly, the number of vehicles available at station $i$ can be updated recursively based on vehicle departures due to transporting passengers $x_i[t-1]$ or rebalancing $y_i[t-1]$, and vehicles arriving from other nodes. Thus, the update equation for each node is

\begin{align}
   &p_i[t]=p_i[t-1]-x_i[t-1]+b_i[t], \\
   &v_i[t]=v_i[t-1]-x_i[t-1]-y_i[t-1]+\sum_{j\neq i} (x_{ji}[t-t_{ji}]+y_{ji}[t-t_{ji}]).
\end{align}

The goal is to find the set of rebalancing vehicles $\{y_{ij}[t]\}_{i,j\in\mathcal{N}}$ such that the average passenger waiting time and total Vehicle Miles Traveled (VMT) due to rebalancing are optimized. The following subsection will address the optimization problem more precisely.

\subsection{SMDP formulation}
\label{sec:MDPFormulation}
The optimization problem of rebalancing vehicles is modeled as a semi Markov Decision Process (SMDP), which is defined by a tuple $(\mathcal{S}, \mathcal{A}, \mathcal{P}, R, \gamma)$. $\mathcal{S}$ and $\mathcal{A}$ represent the state and action space. $\mathcal{P}$ is the transition probability matrix. $R: \mathcal{S}\times\mathcal{A}\to\mathbb{R}$ is the reward function and $\gamma$ is the discount factor. We explain the details as follows:

\begin{itemize}

    \item \textbf{State Space $\mathcal{S}$:} 
    
    The state $s[t]\in\mathcal{S}$ is defined as the information set including passenger queue length $p[t]\triangleq\{p_i[t]\}_{i=1}^n$, vehicle queue length $v[t]\triangleq\{v_i[t]\}_{i=1}^n$, and the history of vehicle departures $(x^{t-1},y^{t-1})$ due to transporting passengers and rebalancing, up to the maximum traveling time $t_{\max}$ between any two of the nodes, that is:
    
    \begin{align*}
        x^{t-1} &\triangleq \big\{x_{ij}[\tau]\big\vert i,j\in\mathcal{N};t-t_{\max}\leq\tau\leq t-1 \big\},\\
        y^{t-1} &\triangleq \big\{y_{ij}[\tau]\big\vert i,j\in\mathcal{N};t-t_{\max}\leq\tau\leq t-1 \big\},\\
        s[t] &\triangleq \{p[t],v[t],x^{t-1},y^{t-1}\}\in\mathbb{R}^{2(n+n^2t_{\max})}.
    \end{align*}
    
    \item \textbf{Action Space $\mathcal{A}$:} 

    The action $a[t]\triangleq\{a_{ij}[t]\}_{i,j\in\mathcal{N}}\in\mathcal{A}\subset\mathbb{R}^{n\times n}$ is defined in general as the set of probability distributions of rebalancing vehicles at each node. That is, if the number of vehicles is greater than the number of waiting passengers at station $i$, then the probability of rebalancing the excessive vehicles to node $j$ is $a_{ij}[t]$, where
    
    $$\sum_{j=1}^n a_{ij}[t]=1 \quad \text{ for all } i\in\mathcal{N}.$$
    
    To reduce the size of the action space, a deterministic action is considered where $a_{ij}[t]\in\{0,1\}$. We assume that the vehicles are dispatched to only the $k$ nearest neighbors for the purpose of rebalancing. Denote the set of $k$ nearest neighborhoods for node $i$ as $\textit{Neighbor}(i)$. The action space is thus given by
    
    \begin{align*}
        a_{ij}[t] \in \{0,1\}\quad\text{if } j\in\textit{Neighbor}(i)\cup\{i\}, \text{ and } a_{ij}[t] =0 \quad \text{otherwise}.
    \end{align*}
    
    We apply the rebalancing action every few time steps. For our simulations, the time step of the SMDP is 1 second whereas the rebalancing action is applied every 100 seconds.
    
    \item \textbf{Transition Probability $\mathcal{P}$:} The transition probability is a map $\mathcal P:\mathcal{S}\times\mathcal{A}\times\mathcal{S} \to [0,1]$. Since the arrival of passengers follows a Poisson process (memoryless), the state $s[t+1]$ follows the distribution $\mathcal{P}(s[t+1]|s[t], a[t])$, which depends on the arrival rates $\lambda_{ij}$, travel time $t_{ij}$. This transition probability is induced by the state update equation and the actions introduced above. 
\end{itemize}

In transportation systems, the quality of service for passengers in terms of their waiting time and the operational cost of rebalancing the fleet need to be traded off. As stated in \cite{henao2019impact}, the customer receives shorter waiting time with ride‑hailing apps but ridehailing leads to approximately 83.5\% more VMT than non-ridehailing cases. We explicitly address this trade-off by picking a cost function that is a convex combination of these two quantities.

\begin{itemize}
    \item \textbf{Reward Function $R$:} According to Little's Law \cite{little1961littleslaw}, the passenger waiting time is proportional to the passenger queue length. Thus, passenger queue length $p_i[t]$ is used to compute passenger waiting time in the cost function. The reward function (negative of cost function) is written as a combination of passenger queue length and the distance traveled by empty vehicles with a weight $\alpha$:
    
    \begin{align} 
    \label{equ:reward}
        R(s[t],a[t])=- \sum_{i=1}^n p_i[t]- \alpha \sum_{i=1}^n \sum_{j \in {\textit{Neighbor}(i)}} \max(v_i[t]-p_{i}[t],0) a_{ij}[t] t_{ij} v.
    \end{align}
    
    The first term in the reward function uses $p_i[t]$ to represent the average waiting time. The second term in the reward function calculates EVMT for rebalancing with $\alpha$ presents the preference of the service providers between the passenger waiting time and EVMT. 
    
    In the second term, $\max(v_i[t]-p_{i}[t],0)$ is the number of vehicles available at station $i$ after serving the local demand at station $i$. The rebalancing distance for vehicles traveling from station $i$ to $j$ is then determined by $a_{ij}[t]t_{ij}v$ where $t_{ij}v$ is the distance between station $i$ and $j$, which is calculated by multiplying the fixed trip time $t_{ij}$ and the average speed $v$ of the vehicles. Summing over $i\in\mathcal{N}$ and $j \in {\textit{Neighbor}(i)}$ gives the total empty vehicle miles traveled. 
    
    \item \textbf{Discount factor $\gamma$:} The discount factor $\gamma$ is set to be 0.99 with respect to the time step $t=1$ second for a infinite horizon MDP Problem.
\end{itemize}

Given a policy $\pi:\mathcal{S}\times\mathcal{A}\to[0, 1]$, the corresponding objective function of the optimization problem can be then written as 

\begin{align} \label{equ:ProblemObjective}
    J(\pi)\triangleq\mathbb{E}_{s_0}^\pi\left[\sum_{t=0}^{\infty} \gamma^t R(s[100t],a[100t])\right],
\end{align}

\noindent
where $s_0$ is the initial state, $a[100t]\sim\pi(\cdot|s[100t])$ and $a[q] = 0$ for $q\neq 100t$, and $s[t+1]\sim\mathbb{P}(\cdot|s[t], a[t])$. The optimal dispatch policy $\pi^*$ maximizes the discounted reward $J(\pi)$. We again emphasize that the rebalancing action is applied every 100 seconds in our simulator. Hence, this is an SMDP problem.

As we can observe, the objective function and the state update equation is very complicated, due to which solving this analytically is seemingly difficult. As a result, we adopt an RL approach to devise an approximately optimal rebalancing policy. We next review the theory of proximal policy optimization with a neural network as a function approximator, which is the primary RL tool used to solve this problem in our paper. There are, of course, other RL algorithms with other function approximators that can be used to solve this problem.

\begin{table}[]
\centering
\begin{tabular}{@{}ll@{}}
\toprule
Notations & Meaning \\ \midrule
$\mathcal{N}$ & station set\\
$\mathcal{E}$ & topology set between stations\\
$\mathcal{S}$ & state space\\
$\mathcal{A}$ & action space\\
$\mathcal{P}$ & state transition probability\\
$\gamma$ & discount factor\\
$\pi(a|s, \theta), \pi_{\theta}$ & parameterized policy with parameter $\theta$\\
$R(s, a)$ &  reward given state $s$ and action $a$\\
$V_{\pi}(s)$ &  state value function under $\pi$\\
$Q_{\pi}(s, a)$ &  action value function under $\pi$\\
$\eta(\pi)$ &  expected reward under $\pi$\\
$A_{\pi}(s, a)$ &  advantage function under $\pi$\\
$L_{\pi}(\tilde{\pi})$ &  expected advantage function under $\pi$\\
$M_{\pi}(\tilde{\pi})$ &  surrogate objective\\
$D_{KL}(\pi, \tilde{\pi})$ &  Kullback–Leibler divergence \\
  &         \\ \bottomrule
\end{tabular}
\caption{The notations used throughout the paper.}
\label{tab: Notations}
\end{table}

\section{Solution Approach: Proximal Policy Optimization} \label{sec:algorithm}

In this section, we discuss the proximal policy optimization algorithm (PPO) adopted for the taxi rebalancing problem. PPO is the state-of-the-art policy optimization algorithm that has gained impressive success in continuous state space control problems \cite{hamalainen2018ppo}. In general, policy optimization algorithms use a parameterized stochastic policy function $\pi(a|s, \theta)$ and formulates a general nonlinear optimization problem of the expected discounted reward over that parameterized class of policies. In practice, the parameterized policy is often represented in the form of a neural network. Therefore, it is restricted to a small class of functions $\Pi = \{\pi(a|s,\theta):\theta\in\Theta\}$ where $\Theta$ is the neural network parameter space. We use $\pi_\theta$ to denote $\pi(a|s, \theta)$, and $J(\theta)$ to denote

$$J(\theta)\triangleq\mathbb{E}_{s_0}^{\pi_\theta}\left[\sum_{t=0}^{\infty} \gamma^t R(s[100t],a[100t])\right].$$

The PPO algorithm consists of two main components: the surrogate objective for expected policy reward; and the policy neural network and value function neural network with different or shared parameters. PPO outperforms the other policy gradient methods and Q-learning based methods, especially in continuous control tasks, due to the following advantages. PPO tends to monotonically improve the policy in terms of the expected discounted reward \cite{schulman2017proximal}. The magnitude of the improvement can be controlled by a tunable parameter, which increases the overall stability and robustness of the algorithm.

\subsection{Surrogate Objective}

Though directly optimizing $J(\theta)$ has been used in early policy optimization algorithms, such as the REINFORCE \cite{williams1992simple} and policy gradient \cite{sutton2000policy}, they generally suffer from sample inefficiency due to inability to reuse samples for gradient estimation. Instead, PPO considers the problem of maximizing policy improvement. The difference in expected reward between policies can be captured by the advantage $A_{\theta}(s, a)$ as shown in \cite{schulman2015trust}:
\begin{equation}
\label{eqn:PolicyImprovement}
    J(\theta') -J(\theta) =  \mathbb{E}_{s_0}^{\pi_{\theta'}}\left[\sum_{t=1}^{\infty} \gamma^t A_{\theta}(s[t],a[t])\right].
\end{equation}
The theoretical results proposed by Schulman et al. \cite{schulman2015trust} provide an explicit lower bound on the expected reward of a new policy $J(\theta')$ relative to the expected reward of the old policy $J(\theta)$. The lower bound uses an local approximation to the right side of \eqref{eqn:PolicyImprovement} which can be easily estimated from samples. The original policy improvement maximization be converted to an equivalent problem of maximizing its lower bound. Through further derivation and relaxation shown in \cite{schulman2015trust}, one can arrive at the surrogate objective in \eqref{eqn: SurrogateLoss}, which is simpler to compute and optimize in practice. 

\subsubsection{Lower Bound Maximization}
Consider the expected discounted reward $J(\theta)$, at each iteration, PPO algorithm tries to find an updated policy parameterized by $\theta'$ given the current $\theta$ such that
\begin{equation}
    \pi_{\theta'} = \underset{\theta'\in\Theta}{\argmax}\ J(\theta') - J(\theta).
\end{equation}
\noindent
However, this optimization problem can be difficult to solve due to the presence of $\pi_{\theta'}$ in the expectation in \eqref{eqn:PolicyImprovement}. Instead, the PPO algorithm considers a surrogate objective \cite{schulman2017proximal} which locally forms a lower bound for the improvement in expected reward. Before introducing the definition of the surrogate objective, let us first define the expected advantage function of policy $\pi_{\theta'}$ over policy $\pi_\theta$.
\begin{definition}
    The expected advantage function $L_{\theta}(\theta')$ of policy $\pi_{\theta'}$ over policy $\pi_\theta$ is defined as
    \begin{equation}
        L_{\theta}(\theta') \triangleq \mathbb{E}_{s\sim \rho_\theta, a\sim \pi_\theta} \left [\frac{\pi_{\theta'}(a|s)}{\pi_\theta(a|s)} A_{\theta}(s, a) \right],
    \end{equation}
    where $\rho_\theta$ is the state visitation frequency under $\pi_\theta$.
\end{definition}
In practice, advantage function $A_{\theta}(s, a)$ is replaced by a generalized advantage estimator (GAE) \cite{schulman2015high} or a fixed horizon estimator \cite{mnih2016asynchronous}. The approximated advantage function is denoted by $\hat{A}_\theta$. Let $D_{KL}(\cdot || \cdot)$ be the KL-divergence between two probability distributions. The following defines the surrogate objective used in PPO:

\begin{definition}
\label{dfn: SurrogateObjective}
    Let $C = \frac{4\epsilon\gamma}{(1-\gamma)^2}$, and  $\epsilon = \underset{s, a}{\max}\ |A_{\theta}(s, a)|$. Define the surrogate objective $M_{\theta}(\theta')$ such that:
    \begin{equation}
    \label{eqn: SurrogateLoss}
        M_{\theta}(\theta') = L_{\theta}(\theta') - CD_{KL}^{\max}(\theta, \theta'),
    \end{equation}\\
    where,
    \begin{equation*}
        D_{KL}^{\max}(\theta, \theta') = \underset{s\in\ALP S}{\max}\  D_{KL}(\pi_\theta(\cdot|s)||\pi_{\theta'}(\cdot|s)).
    \end{equation*}
\end{definition}

In the original results \cite{kakade2002approximately}, the total variation divergence was used in place of KL-divergence in \eqref{eqn: SurrogateLoss}. By exploiting the fact that $D_{TV}(p||q)^2 \leq D_{KL}(p||q)$ \cite{pollard2000asymptopia}, Schulmean et al. \cite{schulman2015trust} derived the lower bound for generalized stochastic policy as in Definition \ref{dfn: SurrogateObjective}, which is a relaxation of the lower bound in \cite{kakade2002approximately}. Hence, we have the lower bound holds as shown in Theorem \ref{thm: LowerBound}, with equality holds if and only if $\theta' = \theta$.

\begin{theorem}[\cite{schulman2015trust}]
\label{thm: LowerBound}
    The following bound holds:
    \begin{equation}
    \label{eqn: LowerBound}
        J(\theta')-J(\theta) \geq L_\theta(\theta') - CD_{KL}^{max}(\theta, \theta').
    \end{equation}
\end{theorem}

Given the current policy $\pi_\theta$ at each iteration, PPO algorithm obtains the updated policy $\pi_{\theta'}$ that maximizes the surrogate objective $M_\theta(\theta')$, which is defined in \eqref{eqn: SurrogateLoss}. According to \eqref{eqn: LowerBound}, the PPO algorithm is guaranteed to generate a monotonically improving sequence of policies.

\subsubsection{Trust region constraint}
In the previous subsection, we showed that updating policy $\pi$ parameterized by $\theta$ to $\theta'$ using PPO algorithm is equivalent to

\begin{equation} \label{eqn: PenaltyObjective}
    \underset{\theta'\in\Theta}{\maximize}\ \left[ L_{\theta}(\theta') -CD_{KL}^{\max}(\theta', \theta)\right].
\end{equation}

This subsection shows how to solve this by rewriting it to its equivalent trust region formulation. This transformation is used mainly for two practical reasons \cite{schulman2015trust}: 1) if the formulation in \eqref{eqn: PenaltyObjective} and the KL-divergence coefficient $C$ takes the value given in Definition \ref{dfn: SurrogateObjective} then the step sizes would be very small; 2) the $\delta$ parameter in \eqref{eqn: TrustRegionConstraint} is easier to tune as it directly represents the maximum allowed change from $\theta$ to $\theta'$.

\begin{align}
    \label{eqn: ConstraintObjective}
    \underset{\theta'\in\Theta}{\maximize} \quad & L_{\theta}(\theta')\\
    \label{eqn: TrustRegionConstraint}
    \subjectto \quad & D_{KL}^{\max}(\theta, \theta') \leq \delta.
\end{align}

However, \cite{schulman2017proximal} proposed a third formulation, the clipped surrogate objective $L^{CLIP}$, which is much simpler to implement and tune. It even outperforms other formulations in simulation results. Since the maximization in Equation $\eqref{eqn: ConstraintObjective}$ contains a probability ratio between $\theta'$ and $\theta$ which reflects the size of the policy update, we can restrict the probability ratio to be within some neighborhood around 1 and clip away the large updates. The objective function in Algorithm \ref{algo: Algorithm} uses the clipped surrogate objective. The details can be found in \cite{schulman2017proximal}.

\subsection{The Complete Algorithm}

So far, we have discussed the surrogate objective and how it forms the clipped objective $L^{CLIP}$ used in the algorithm. Now, we can construct the complete algorithm as shown in Algorithm \ref{algo: Algorithm}. $\theta_1, \theta_2 \in \Theta$ are the parameters for policy network and value function network respectively. They are initialized arbitrarily as well as the initial state $s_0$. For each iteration, the algorithm samples the environment up to $T$ timesteps using the policy from the previous iteration. The advantage $\hat{A}_t$ at timestep $t$ is then estimated using the samples and the approximated value function $V(s,\ \theta_{2})$ \cite{mnih2016asynchronous}. After that, $K$ gradient updates (epochs) are performed by drawing minibatches of samples. The gradient of $L^{CLIP}$ w.r.t $\theta_1$ and the gradient of value function loss w.r.t $\theta_2$ are obtained using the automatic differentiation engine in the Tensorflow framework. 

\begin{algorithm}
\SetAlgoLined
\textbf{INITIALIZE:} $\theta_1, \theta_2, \pi_{0}, s_{0}$\\
$a_{0} \sim \pi_{0}(a|s_{0})$\\
\For{iteration: i=1,2,...}{
    Run policy $\pi_{i-1}$ for $T$ timesteps \\
    Collect samples $S_{i} = \{(s_1, a_1, r_1), (s_2, a_2, r_2), ..., (s_{T-1}, a_{T-1}, r_{T-1}), (s_T, a_T, r_T)\}$\\
    Compute advantage estimates $\hat{A}_{1}, \hat{A}_{2}, ..., \hat{A}_{T-1}$\\
    \For{epoch: k=1,2,...,K}{
        Draw a minibatch $B_{k} \subset S_{i}$ from collected samples in this iteration\\
        Compute gradient $g_1 \leftarrow \nabla_{\theta_1} L^{CLIP}_{\theta_1}$\\
        Compute gradient $g_2 \leftarrow \nabla_{\theta_2} (V(s_t,\ \theta_2^i) - r_t - V(s_{2},\ \theta_{2}^{i-1}))^2$\\
        $\theta_1 \leftarrow \theta_1 + \alpha g_1$\\
        $\theta_2 \leftarrow \theta_2 + \alpha g_2$
    }
}
\textbf{OUTPUT:} $\theta_1,\theta_2$\\
\caption{Proximal Policy Optimization \cite{schulman2017proximal}}
\label{algo: Algorithm}
\end{algorithm}

\section{Experiment Setup}\label{sec:setup}

In this section, we discuss the details of our simulation setup. We first discuss the procedure adapted for data processing of New York Taxi data to yield arrival rates of passengers. Thereafter, we present the details of the simulator written in Python to carry out the RL implementation. Finally, we review various rebalancing algorithms that have been proposed in the recent literature for benchmarking the performance of the trained policy. 
\subsection{Data Processing} \label{sec:dataproc}
We obtain the data from New York City Taxi and Limousine Commission website \cite{website_nyc}. The data is analyzed to extract the information on the passenger arrival process. We specifically pick the Manhattan borough yellow cab trip records between 8 am to 9 am on weekdays in December 2019. This time window is chosen because we observed that the demand for taxis is high in this time interval and the demand is highly asymmetric. The imbalanced rush hour scenario poses a greater challenge to the rebalancing policy. From our data analysis, we found that the real demand follows Poisson arrival since the interarrival time distribution has a good fit with the exponential distribution. For more details of data processing, we refer the reader to \ref{app:dataprocess}.

\subsection{Simulator} \label{sec:simulator}

\begin{figure}[H]
    \centering
    \includegraphics[width=0.9\textwidth]{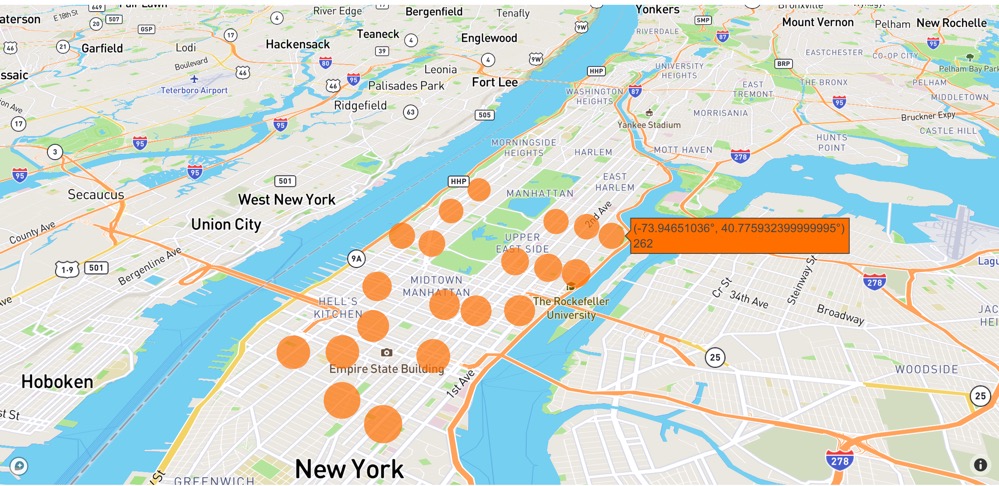}
    \centering
    \caption{20 nodes network in Midtown Manhattan}
    \label{fig:topo_20}
\end{figure}

Our simulator has a stationary passenger arrival process simulated by a Poisson process with arrival rates estimated from the selected data set. The topology of the network is modeled after the geographic taxi zones in the Manhattan borough minus some isolated zones have no ground connection with others. The complete Manhattan taxi zone map can be found in Figure \ref{fig_manhattanZones}. The simulator has 63 zones in total, and one can configure it to use any subset of the zones. Figure \ref{fig:topo_20} shows the topology based on a 20 nodes network around Midtown Manhattan. Our RL simulations are conducted for the demand across these 20 nodes.

The entire process of the simulator can be summarized as four steps: 1) passenger generation; 2) matching vehicles and passengers; 3) dispatch empty vehicles; 4) vehicle arrival. Passengers accumulate at the stations if there are not enough vehicles to pick them up. When a vehicle is at a node with no passenger waiting, the taxi can either choose to wait at the station or be dispatched (rebalanced) to the other nodes with higher demands. The simulator provides performance statistics for comparison between rebalancing policies, which includes the average passenger waiting time, number of rebalancing trips, and rebalancing miles traveled. 

Here, we list the general simulator settings that common for both the PPO algorithm and the benchmarks. The simulator updates the passenger queues, vehicle distribution, and other system statuses every second. The rebalancing policies are executed at a fixed frequency of every 100 seconds, which is also known as the rebalancing interval. For the PPO algorithm, this is also the frequency at which the state is observed and action is applied (one MDP timestep). Also, each episode is defined as 10 hours of simulation from a given initial state. 

To gain more control over the tuning process, we made the following minor changes to the system model. Recall that the control action defined in Section \ref{sec:MDPFormulation} can only take value from $\{0, 1\}$. In the simulator, we added a dispatch ratio ($dpr$) as a tunable parameter that controls the proportion of the available vehicles in a station that can be dispatched to the chosen station.

{}

\begin{table}[]
\centering
\resizebox{\textwidth}{!}{%
\begin{tabular}{@{}lrllllllllllllllllllll@{}}
                            & \multicolumn{1}{l}{}     & \multicolumn{20}{c}{Destination}                                                                                                          \\ \cmidrule(l){2-22} 
\multicolumn{1}{l|}{}       & \multicolumn{1}{r|}{}    & 48   & 68   & 100  & 107  & 140  & 141  & 142  & 143  & 161  & 162  & 170  & 186  & 229  & 234  & 236  & 237  & 238  & 239  & 262  & 263  \\ \cmidrule(l){2-22} 
\multicolumn{1}{l|}{}       & \multicolumn{1}{r|}{48}  & 0.00 & 0.74 & 0.18 & 0.63 & 2.43 & 2.09 & 0.61 & 0.30 & 0.84 & 1.21 & 0.83 & 0.32 & 1.71 & 0.42 & 2.29 & 1.68 & 1.29 & 0.88 & 3.01 & 2.70 \\
\multicolumn{1}{l|}{}       & \multicolumn{1}{r|}{68}  & 0.74 & 0.00 & 0.78 & 1.12 & 3.14 & 2.81 & 1.36 & 1.00 & 1.55 & 1.91 & 1.48 & 0.52 & 2.41 & 0.67 & 3.03 & 2.40 & 2.03 & 1.62 & 3.73 & 3.43 \\
\multicolumn{1}{l|}{}       & \multicolumn{1}{r|}{100} & 0.18 & 0.78 & 0.00 & 0.46 & 2.36 & 2.03 & 0.63 & 0.44 & 0.77 & 1.14 & 0.72 & 0.27 & 1.63 & 0.28 & 2.25 & 1.63 & 1.31 & 0.91 & 2.95 & 2.65 \\
\multicolumn{1}{l|}{}       & \multicolumn{1}{r|}{107} & 0.63 & 1.12 & 0.46 & 0.00 & 2.10 & 1.78 & 0.72 & 0.79 & 0.60 & 0.89 & 0.44 & 0.62 & 1.36 & 0.45 & 2.05 & 1.41 & 1.29 & 0.97 & 2.70 & 2.42 \\
\multicolumn{1}{l|}{}       & \multicolumn{1}{r|}{140} & 2.43 & 3.14 & 2.36 & 2.10 & 0.00 & 0.34 & 1.86 & 2.28 & 1.59 & 1.23 & 1.68 & 2.62 & 0.74 & 2.51 & 0.33 & 0.76 & 1.36 & 1.69 & 0.60 & 0.36 \\
\multicolumn{1}{l|}{}       & \multicolumn{1}{r|}{141} & 2.09 & 2.81 & 2.03 & 1.78 & 0.34 & 0.00 & 1.52 & 1.94 & 1.26 & 0.90 & 1.35 & 2.29 & 0.43 & 2.19 & 0.31 & 0.42 & 1.04 & 1.35 & 0.92 & 0.64 \\
\multicolumn{1}{l|}{}       & \multicolumn{1}{r|}{142} & 0.61 & 1.36 & 0.63 & 0.72 & 1.86 & 1.52 & 0.00 & 0.42 & 0.40 & 0.71 & 0.54 & 0.89 & 1.18 & 0.88 & 1.70 & 1.10 & 0.68 & 0.28 & 2.42 & 2.11 \\
\multicolumn{1}{l|}{}       & \multicolumn{1}{r|}{143} & 0.30 & 1.00 & 0.44 & 0.79 & 2.28 & 1.94 & 0.42 & 0.00 & 0.77 & 1.12 & 0.83 & 0.62 & 1.60 & 0.71 & 2.12 & 1.53 & 1.05 & 0.64 & 2.84 & 2.53 \\
\multicolumn{1}{l|}{}       & \multicolumn{1}{r|}{161} & 0.84 & 1.55 & 0.77 & 0.60 & 1.59 & 1.26 & 0.40 & 0.77 & 0.00 & 0.37 & 0.20 & 1.03 & 0.87 & 0.94 & 1.49 & 0.86 & 0.72 & 0.50 & 2.18 & 1.89 \\
\multicolumn{1}{l|}{origin} & \multicolumn{1}{r|}{162} & 1.21 & 1.91 & 1.14 & 0.89 & 1.23 & 0.90 & 0.71 & 1.12 & 0.37 & 0.00 & 0.46 & 1.40 & 0.50 & 1.29 & 1.15 & 0.52 & 0.67 & 0.68 & 1.82 & 1.53 \\
\multicolumn{1}{l|}{}       & \multicolumn{1}{r|}{170} & 0.83 & 1.48 & 0.72 & 0.44 & 1.68 & 1.35 & 0.54 & 0.83 & 0.20 & 0.46 & 0.00 & 0.96 & 0.94 & 0.84 & 1.61 & 0.97 & 0.92 & 0.69 & 2.27 & 1.99 \\
\multicolumn{1}{l|}{}       & \multicolumn{1}{r|}{186} & 0.32 & 0.52 & 0.27 & 0.62 & 2.62 & 2.29 & 0.89 & 0.62 & 1.03 & 1.40 & 0.96 & 0.00 & 1.89 & 0.21 & 2.52 & 1.89 & 1.57 & 1.17 & 3.22 & 2.92 \\
\multicolumn{1}{l|}{}       & \multicolumn{1}{r|}{229} & 1.71 & 2.41 & 1.63 & 1.36 & 0.74 & 0.43 & 1.18 & 1.60 & 0.87 & 0.50 & 0.94 & 1.89 & 0.00 & 1.78 & 0.72 & 0.23 & 0.86 & 1.07 & 1.34 & 1.06 \\
\multicolumn{1}{l|}{}       & \multicolumn{1}{r|}{234} & 0.42 & 0.67 & 0.28 & 0.45 & 2.51 & 2.19 & 0.88 & 0.71 & 0.94 & 1.29 & 0.84 & 0.21 & 1.78 & 0.00 & 2.43 & 1.80 & 1.55 & 1.17 & 3.11 & 2.82 \\
\multicolumn{1}{l|}{}       & \multicolumn{1}{r|}{236} & 2.29 & 3.03 & 2.25 & 2.05 & 0.33 & 0.31 & 1.70 & 2.12 & 1.49 & 1.15 & 1.61 & 2.52 & 0.72 & 2.43 & 0.00 & 0.64 & 1.13 & 1.50 & 0.73 & 0.42 \\
\multicolumn{1}{l|}{}       & \multicolumn{1}{r|}{237} & 1.68 & 2.40 & 1.63 & 1.41 & 0.76 & 0.42 & 1.10 & 1.53 & 0.86 & 0.52 & 0.97 & 1.89 & 0.23 & 1.80 & 0.64 & 0.00 & 0.68 & 0.94 & 1.33 & 1.03 \\
\multicolumn{1}{l|}{}       & \multicolumn{1}{r|}{238} & 1.29 & 2.03 & 1.31 & 1.29 & 1.36 & 1.04 & 0.68 & 1.05 & 0.72 & 0.67 & 0.92 & 1.57 & 0.86 & 1.55 & 1.13 & 0.68 & 0.00 & 0.41 & 1.86 & 1.54 \\
\multicolumn{1}{l|}{}       & \multicolumn{1}{r|}{239} & 0.88 & 1.62 & 0.91 & 0.97 & 1.69 & 1.35 & 0.28 & 0.64 & 0.50 & 0.68 & 0.69 & 1.17 & 1.07 & 1.17 & 1.50 & 0.94 & 0.41 & 0.00 & 2.22 & 1.91 \\
\multicolumn{1}{l|}{}       & \multicolumn{1}{r|}{262} & 3.01 & 3.73 & 2.95 & 2.70 & 0.60 & 0.92 & 2.42 & 2.84 & 2.18 & 1.82 & 2.27 & 3.22 & 1.34 & 3.11 & 0.73 & 1.33 & 1.86 & 2.22 & 0.00 & 0.32 \\
\multicolumn{1}{l|}{}       & \multicolumn{1}{r|}{263} & 2.70 & 3.43 & 2.65 & 2.42 & 0.36 & 0.64 & 2.11 & 2.53 & 1.89 & 1.53 & 1.99 & 2.92 & 1.06 & 2.82 & 0.42 & 1.03 & 1.54 & 1.91 & 0.32 & 0.00 \\ \cmidrule(l){3-22} 
\end{tabular}%
}
\caption{Distances between every pair of nodes in the 20 node network in miles. We assume that the average velocity is 10 miles per hour on the road. Thus, $t_{\max} = 22$ mins.}
\label{tab: TravelDistance}
\end{table}

\subsection{PPO algorithm settings}

A summary of the configurations for the PPO algorithm is shown in Table \ref{tab: hyperparameter}. We set the discount factor $\gamma = 0.99$. The neural network has an input vector with length equal the number of taxi stations and two hidden layers each with 256 units and tanh nonlinearity. The initial weights for the neural network are drawn from a uniform distribution. The optimizer of choice is the Adam algorithm \cite{kingma2014adam}. The value function and the policy use separate networks, and each uses a learning rate of $3\times 10^{-4}$. The output of the policy network is a vector with length equal to the number of taxi stations, and the output of the value function network is a scalar. The training process takes 2000 iterations. Within one iteration, the PPO algorithm uses a fixed policy to sample state, action, and reward data from the simulator. Each iteration consists of 4000 timesteps, and one timestep equals one rebalancing interval mentioned in the above section which is 100 seconds in simulation time. The optimizer performs 30 gradient ascent steps on the samples collected in each iteration. The sample sets are randomly picked to form a minibatch of 128 samples on which the gradient estimation is performed. The training process is executed using Ray \cite{liaw2018tune}, a distributed training framework, which allows us to run multiple sampling agent simultaneously. After training, the trained policy is applied to 10 test episodes. The average performance metrics of the test episodes are used for comparison with the benchmarks.

\begin{table}[h!]
\centering
    \begin{tabular}{l l}
    \hline
    Parameter &Value\\
    \hline
    optimizer &Adam\\
    learning rate &$3\times 10^{-4}$\\
    discount($\gamma$) &0.99\\
    training batch size &4000\\
    SGD minibatch size &128\\
    number of hidden layers &2\\
    activation function &tanh\\
    number of hidden units per layer &256\\
    number of samples per minibatch &256\\
    gradient steps &30\\
    \hline
    \end{tabular}
\caption{PPO Hyperparameters used in the training.}
\label{tab: hyperparameter}
\end{table}

\subsection{Benchmarks}
We compare our approach with the following existing algorithms and heuristics.

\subsubsection{MaxWeight}

MaxWeight policy is an online scheduling algorithm and has been used extensively across a wide range of applications. For the ridehailing system, the properties of MaxWeight policy are studied in \cite{Banerjee2018}. The policy works as follows: If a passenger arrives at station $i\in \mathcal{N}$ and the number of vehicle at that station is zero $v_i[t]=0$, then the algorithm will dispatch an empty vehicle from the station

\begin{align*}
    j^* = \underset{j \in Neighbor(i)}{\argmax} \{v_1[t], \ldots, v_n[t]  \},
\end{align*}
and match this vehicle with the passenger. The dispatch continues until all passengers get matched. There is no dispatch decision if the passengers do not queue up.


\subsubsection{Mirror BackPressure}

This policy was designed in a recent work \cite{Kanoria2019}. The policy assigns a score to each pair of dispatching origin $j$ and destination $i$, which depends on the number of vehicle $v_j[t]$ and the distance between two nodes $d_{ji}$.  It will only dispatch a vehicle from origin to destination if the score $\textit{score}_{ji}[t]$ is positive. 
 Let the score defined as $\textit{score}_{ji}[t]=f(v_j[t])-d_{ji}$ and the set of score: $ \textit{score}_{i}[t]= \{\textit{score}_{1i}[t], \ldots, \textit{score}_{ni}[t]\}$  the algorithm will dispatch an empty vehicle to node $i$ from station $j^*$, which is defined as
 
\begin{align*}
    j^* =
    \begin{cases}
        \underset{j \in Neighbor(i)}{\argmax} \textit{score}_{i}[t]  & \underset{j \in Neighbor(i)} \max \textit{score}_{i}[t] >0 \\
        \emptyset  &\underset{j \in Neighbor(i)} \max \textit{score}_{i}[t] \leq 0
    \end{cases}.
\end{align*}


\subsubsection{Proportional}

This heuristic will dispatch vehicles to nodes in its neighborhood proportional to the passenger queue length $p_i[t]$ at that node. No vehicle is dispatched when the sum of passenger queue length is zero around the neighborhood. If there are extra empty vehicles at station $j$, that is, $v_j[t]-p_j[t] >0$, then the number of empty vehicles dispatched from node $j$ to node $i$ is as follows:

\begin{align*}
    y_{ji}[t]=(v_j[t]-p_j[t]) \frac{p_i[t]}{\sum_{i \in Neighbor(j)} p_i[t]}.
\end{align*}

\subsection{CostSensitive}
This real-time rebalancing policy is proposed in \cite{zhang2016control}, where the number of empty vehicle dispatched every time is decided by solving a linear program. Let $M$ be the total number of vehicle in the system and the number of excess vehicle is defined as $M-\sum_{j=1}^n \max \{ p_i[t]-v_i[t], 0\}$. Through rebalancing, this algorithm wishes to distribute excess vehicles evenly into $n$ stations, i.e. $v_i^d[t]= \floor{(M-\sum_{j=1}^n \max \{ p_i[t]-v_i[t], 0\})/n}$.
Accordingly, the number of rebalancing vehicles from station $i$ to $j$, $y_{ij}[t]$, can be found through following linear integer programming, where $t_{ij}$ is the travel time.

\begin{align*}
    \underset{y_{ij}}{\minimize} \quad &\sum_{i,j} t_{ij} y_{ij} \\
    \subjectto &v_i[t] + \sum_{j \neq i} (y_{ji}-y_{ij}) \geq v_i^d[t], \quad \forall i \in \mathcal{N} \\
\end{align*}

We next show through numerical simulation that the rebalancing policy obtained through PPO outperforms the rebalancing algorithms reviewed above.

\begin{figure*}[!h]
    \centering
    \begin{subfigure}[!h]{0.48\textwidth}
        \centering
        \includegraphics[width=\textwidth]{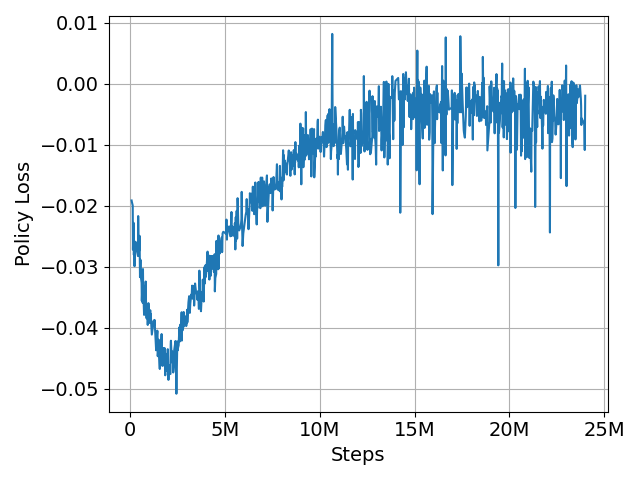}
    \end{subfigure}
    \hfill
    \begin{subfigure}[!h]{0.48\textwidth}
        \centering
        \includegraphics[width=\textwidth]{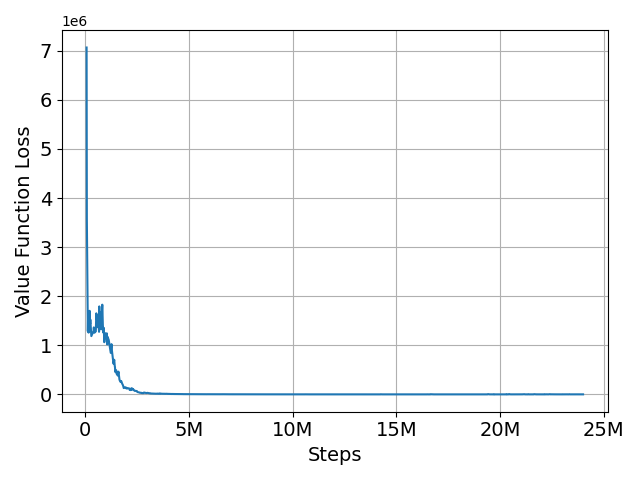}
    \end{subfigure}
     \hfill
     \begin{subfigure}[!h]{0.48\textwidth}
        \centering
        \includegraphics[width=\textwidth]{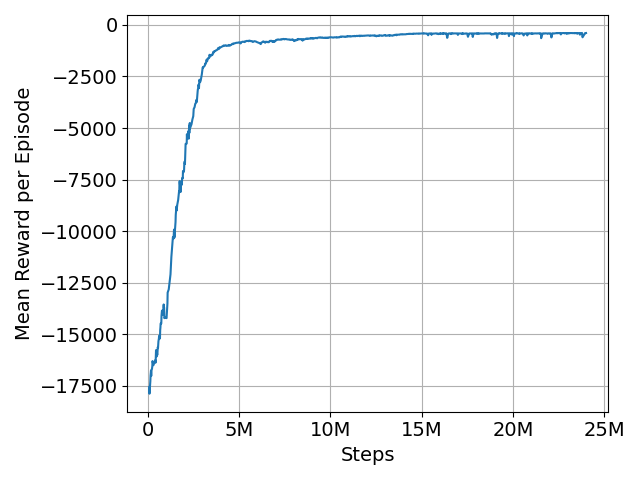}
    \end{subfigure}
    \caption{Policy Loss (Left), Value Function Loss (Right), and accumulated Reward per episode (Below)}
    \label{fig:training}
\end{figure*}

\section{Results and Analysis}\label{sec:results}
In this section, we present the results of our algorithm and compare it to the benchmarks mentioned in the last section. 

\subsection{Training loss}
The policy loss is defined as the negative surrogate objective \eqref{eqn: SurrogateLoss}, defined in Section \ref{sec:algorithm}. The policy loss should approach zero from the bottom as policy gets close to the approximately optimal policy within the policy approximating class. The value function loss is the mean square error between the neural network output and the target value. 
As shown in Figure \ref{fig:training}, the policy converges to a stable policy as the improvement is diminishing, and the value function neural network converges as the loss goes to zero. In the meantime, the reward function (defined in Equation \ref{equ:reward}) increases as training progresses and converges to a value with a small negative value, which indicates negative residual cost when the policy is near-optimal. 

\begin{sidewaysfigure}
    \centering
    \includegraphics[trim={0.2cm 3.5cm 0.2cm 0.1cm},clip,width=1\textwidth]{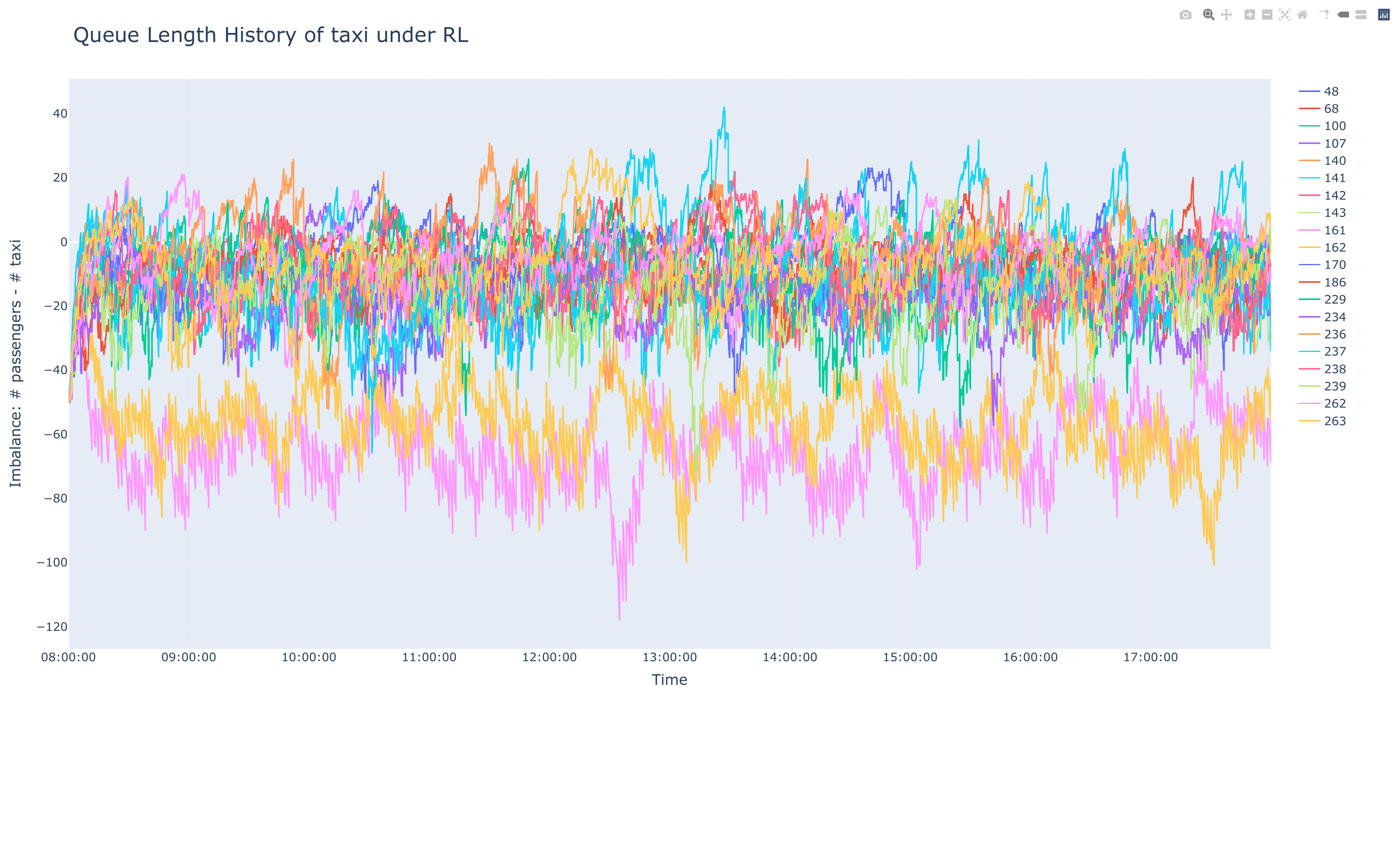}
    \centering
    \caption{Sample path under the rebalancing algorithm}
    \label{fig:imbalance}
\end{sidewaysfigure}

Figure \ref{fig:imbalance} shows one sample path of the rebalancing algorithm, where the Y-axis is the imbalance of each station, i.e. $p_i[t]-v_i[t]$ for station $i \in \mathcal{N}$ and X-axis is time.

\subsection{Performance Comparison}
We conduct experiments on two scenarios with 1000 and 600 fleet sizes. All the vehicles are equally distributed at the beginning of the simulation. We set the speed to 10 mph for all vehicles to reflect the real average traffic speed in Manhattan as suggested in \cite{nycmobility2018}. Here, we introduce the metrics used to compare RL based rebalancing policy against other benchmarks. For the test case, the number of total passengers is 46377 within 10 hours when the initial random seed is set to 0, and fixed for all the test cases. The average waiting time is the period between the arrival of a passenger and his/her departure. The cost of waiting is the product of the average waiting time and the total number of passengers. The Empty Vehicles Miles Traveled (EVMT) representing the rebalancing cost is calculated as the sum of the distance traveled by empty vehicles. We provide relative value for the cost of waiting and EVMT by dividing these values by those of the MaxWeight rebalancing policy.

\begin{table}[h!]
\centering
\begin{tabular}{l  c c c c c}
\hline
Algorithm ($\alpha$) & \shortstack{Avg. wait \\time (mins)}& \shortstack{Rebalance \\ trips} & \shortstack{ Avg.\\ EVMT} & \shortstack{Rel. cost \\wait time} & \shortstack{Relative \\EVMT}\\
\hline
PPO($1e^{-2}$)  &0.21 &30603 &1.07 &0.37 &2.14\\
PPO($1e^{-1}$)  &0.30 &24707 &1.07 &0.53 &1.73\\
PPO($1e^0$)  &0.35 &24580 &0.91 &0.62 &1.46\\
\textbf{PPO($1e^1$)}  &\textbf{0.36} &\textbf{23552} &\textbf{0.64} &\textbf{0.64} &\textbf{0.98}\\
PPO($1e^2$)  &2.75 &18437 &0.65 &4.91 &0.78\\
PPO($1e^3$)  &87.47 &4281 &0.45 &156.19 &0.13\\
MaxWeight  &0.56 &12417 &1.23 &1.00 &1.00\\
BackPressure  &0.75 &35362 &0.55 &1.34 &1.27\\
Proportional  &1.96 &36947 &1.37 &3.50 &3.31\\
CostSensitive  &0.25 &37197 &1.4 &0.45 &3.41\\
None &176.75 &0 &0 &315.62 &0\\
\hline
\end{tabular}
\caption{Performance comparison of various rebalancing algorithms with a fleet size of 1000 vehicles. The passenger arrival process is the same in all algorithms (since they are generated from the same initial random seed). The number of passenger arrivals is 46377 over 10 hours period. The total waiting time of MaxWeight is 25971 minutes and EVMT is 15272, which is set as the baseline for relative comparison.}
\label{tab:performance_1000}
\end{table}

\begin{table}[h!]
\centering
\begin{tabular}{l  c c c c c}
\hline
Algorithm ($\alpha$) & \shortstack{Avg. wait \\time (mins)}& \shortstack{Rebalance \\ trips} & \shortstack{ Avg. \\ EVMT} & \shortstack{Rel. cost \\wait time} & \shortstack{Relative \\EVMT}\\
\hline
\textbf{PPO($1e^1$)}  &\textbf{1.79} &\textbf{21220} &\textbf{0.67} &\textbf{0.41} &\textbf{0.90}\\
PPO($1e^2$)  &3.47 &17839 &0.81 &0.81 &0.92\\
PPO($1e^3$)  &93.88 &4304 &0.44 &218.74 &0.12\\
MaxWeight  &4.29 &12989 &1.21 &1.00 &1.00\\
BackPressure  &19.92 &19418 &0.53 &4.64 &0.65\\
Proportional  &3.52 &12854 &1.21 &0.82 &0.99\\
CostSensitive  &154.06 &2266 &1.1 &35.91 &0.16\\
None &196.94 &0 &0 &45.90 &0\\
\hline
\end{tabular}
\caption{Performance comparison of various rebalancing algorithms with the fleet size of 600 vehicles. The number of passenger arrivals is 46377 over 10 hours period. The total waiting time of MaxWeight is 198957 minutes and EVMT is 15716, which is set as the baseline of relative comparison.}
\label{tab:performance_600}
\end{table}

Table \ref{tab:performance_1000} shows the test performance of each algorithm under 1000 vehicles in the system, where $\alpha$ is the preference parameter defined in Equation \ref{equ:reward}. The total waiting time of MaxWeight is 25971 minutes and EVMT is 15272, which is set as the baseline for comparison. PPO($1e^1$) outperforms MaxWeight, BackPressure, and Proportional with less cost of waiting and less EVMT. Compared with CostSensitive, PPO($1e^{-2}$) has a better performance on both the cost of waiting and the EVMT. Therefore, our method is able to beat all the benchmarks when both metrics are considered by appropriately tuning the parameter $\alpha$.

\begin{figure}[h!]
    \centering
    \includegraphics[trim={0.2cm 0.1cm 0.2cm 0.1cm},clip,width=0.7\textwidth]{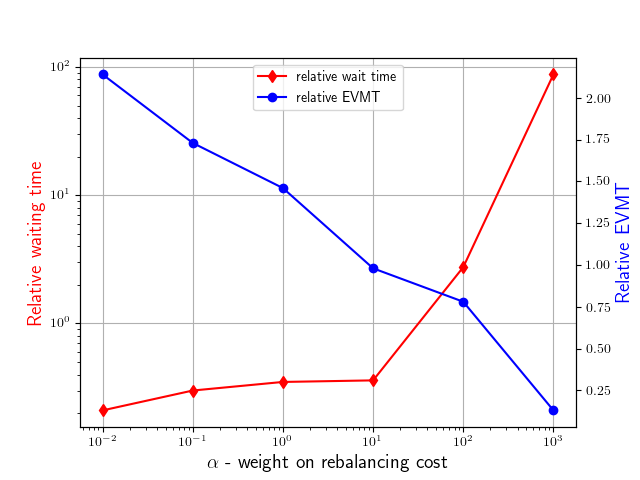}
    \caption{A plot of relative waiting time (in log scale) and relative EVMT versus the weight $\alpha$ for the fleet size of 1000 vehicles. This plot is obtained through the Monte-Carlo simulation to obtain the relative waiting time and relative EVMT. The data is presented in Table \ref{tab:performance_1000}.}
    \label{fig: fleetSize1000}
\end{figure}

As shown in Figure \ref{fig: fleetSize1000}, we can observe that $\alpha$ has the ability to switch the focus of the rebalancing policy between the quality of service and the operational cost. With the decrease of $\alpha$, the algorithm achieves less waiting time at the expense of higher EVMT. This means the service provider can customize the parameter $\alpha$ and tweak the learning objective based on their specific requirements. 

Table \ref{tab:performance_600} shows the performance of RL based rebalancing policies with a smaller fleet size of $600$ vehicles. The number of passenger arrivals is the same with 1000 fleet case and is 46377 over 10 hours period. The total waiting time of MaxWeight is 198957 minutes and EVMT is 15716, which is set as the baseline for relative comparison. PPO($1e^{1}$) achieves the best performance considering waiting time cost and EVMT. Although it has 60\% more rebalancing trips than MaxWeight, it reduces the average empty vehicle miles per rebalancing trip by selecting nearby stations as a rebalancing destination, which leads to 10\% less EVMT compared to MaxWeight.

\section{Conclusion and Future Work}\label{sec:conclusion}
In this paper, we developed an RL-based vehicle rebalancing algorithm for a large-scale ridehailing system. We model the problem within a semi Markov decision problem (SMDP) framework, where passengers queue up at every node until picked up by an empty vehicle. We considered a convex combination of passenger waiting time and empty vehicle miles traveled as the objective function of the SMDP. An intelligent agent makes state-dependent empty vehicle dispatching decisions to maximize the long-term reward, which converges to an approximately optimal solution of the SMDP. The performance of the RL algorithm is compared with other rebalancing algorithms, and we found that RL outperforms other rebalancing algorithms such as MaxWeight and backpressure based rebalancing schemes. The results are encouraging especially when the fleet sizes are small. 

Our future research will address the problem of determining approximately optimal rebalancing schemes through reinforcement learning to also minimize passenger drop probability when passengers are impatient and determining dynamic pricing schemes to curtail demand at nodes with a high arrival rate of passengers. 

\section*{Acknowledgment}
This research is supported in part by Ford Motor Company under the University Alliance Project. We would like to thank Archak Mittal, Richard Twumasi-Boakye, and James Fishelson for informing us about various problems faced by the fleet operators. Results presented in this paper were obtained using the Chameleon testbed supported by the National Science Foundation.

{}
\newpage
\bibliographystyle{elsarticle-num}
\bibliography{final_report}

\newpage

\appendix
\section{Data Processing} \label{app:dataprocess}
The data set used is the monthly records of yellow taxi trips in NYC. The trip records starting from 2009 to 2019 are available from the New York City Taxi and Limousine Commission website (https://www1.nyc.gov/site/tlc/index.page). To reflect the latest traffic condition in NYC, data collected from December 2019 is used as the primary data set to analyze. The total number of data points for December 2019 is roughly 7 million, where each one of the data points includes the following variables:
\begin{itemize}
    \item Vendor ID: This variable indicates which vendor provided the trip data, $1$ for Creative Mobile Tech., LLC and 2 for VeriFone Inc.
    \item Pick-up/Drop-off time: Time when the trip starts/ends.
    \item Passenger: Number of passengers in the trip enter by the driver, ranges from $\{0,1,...,9\}$.
    \item Trip Distance: Miles one trip traveled.
    \item Pick-up/Drop-off location ID: The TLC taxi zone where the trip starts and ends, ranges from $\{1,...,256\}$.
    \item Rate Code ID: The type of rate for the trip, including 1 for standard rate, 2 for JFK, 3 for Newark etc., ranges from $\{1,...,6\}$.
    \item Payment Type: Methods of payment, 1 for credit card, 2 for cash etc., ranges from $\{1,...,6\}$.
    \item Fare Amount: The time-and-distance fare calculated by the meter.
    \item Total Amount: The total amount including fare amount, tips, and other fees.
\end{itemize}
Using this primary data, the following section will cover the main steps used to filter and clean the data, before applying to the simulator.

\subsection{Data Filtering and Cleaning}
\begin{figure}
    \centering
    \begin{subfigure}{0.45\textwidth}
    	\includegraphics[scale=0.2]{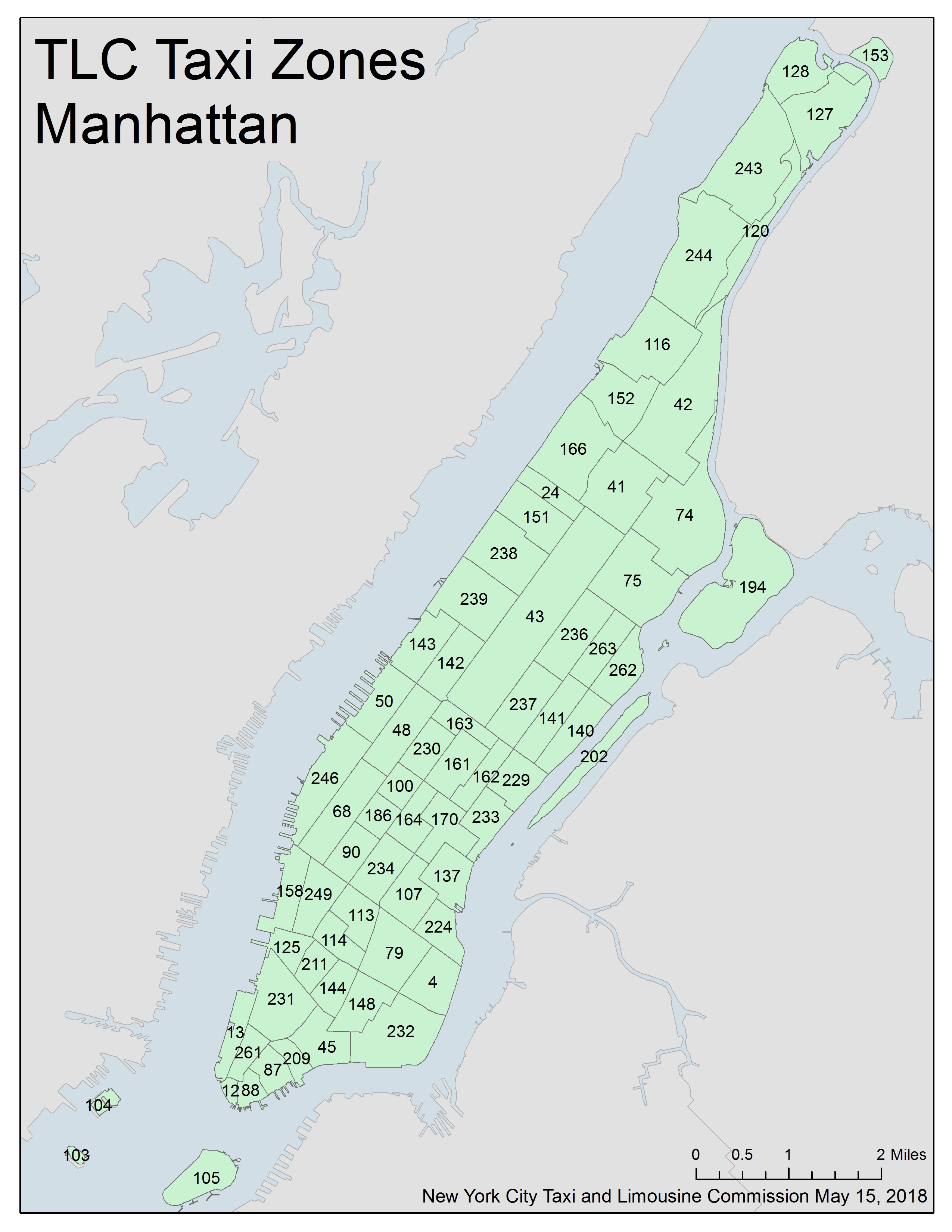}
    	\caption{Zone IDs adopted in original data set}
    \end{subfigure}
    \hskip2em
        \begin{subfigure}{0.45\textwidth}
    	\includegraphics[scale=0.2]{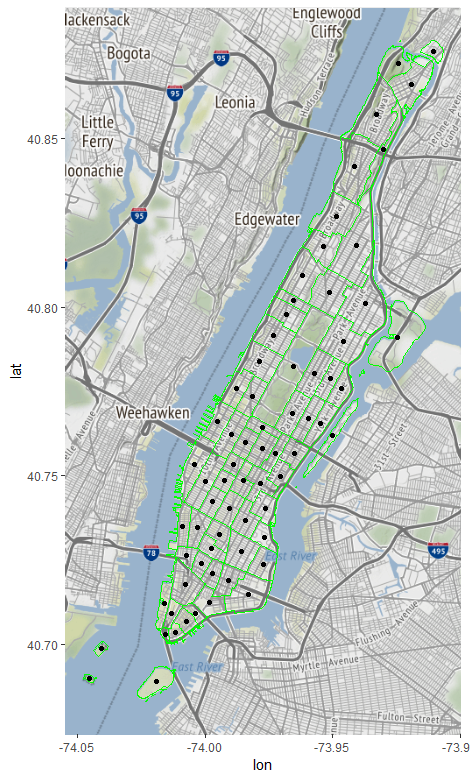}
    	\caption{Zone center calculated in \texttt{R}}
    \end{subfigure}
    \caption{Taxi zones of Manhattan, New York area.}
    \label{fig_manhattanZones}
\end{figure}

Given that the coverage of the data is huge, traffic conditions could be quite different, and thus different models. The simulator considers a one hour window of morning rush time in the downtown area. More specifically, selected trip starts within 8 am to 9 am, and has both pick-up and drop-off points within Manhattan only. The data count is reduced from 6896317 to 294597 points for a one-hour pick-up window and is further reduced to 255183 given trips within Manhattan. After carefully reading the description of each column of the original \texttt{.csv} file, the following criteria are developed to filter out any abnormal data.
\begin{itemize}
    \item Trip time: 60-7200 seconds, calculated from stating and ending time provided. Trip time between 1 minute and 2 hours is required based on the size of the Manhattan area, and estimated driving time of some extreme cases tried from Google Map. The starting and ending time columns will then be deleted.
    \item Trip distance: 0.1-20 miles, existing variable. This range is selected using a similar approach as trip time. Some extremely long trip distance as well as negative trip distance appeared in the dataset and is considered not valid.
    \item Fare amount, total amount: positive value, existing variable. Zero and negative values were observed.
    \item Rate Code ID: 1, existing variable. Trips with standard rate are accepted, other types of trips such as voided or disputed trips are excluded.
    \item Passenger count, 1-6, existing variable. Exclude trips with 0 or \texttt{NA} passenger count.
    \item MTA tax, congestion surcharge, improvement surcharge: a fixed amount, existing variable. Extras are listed on the NYC website. Trip with values other than the mandated will be dropped.
    \item Day of the week: 0-6, calculated based on the trip dates, new variable as a factor.
    \item Latitude and longitude of zone center, calculated based on the zone border using \texttt{gCentroid} in \texttt{rgeos} package. Zone border file can be downloaded from the NYC website and further process can be done using \texttt{rgdal} package. In the original data set, only zone IDs are provided to locate pick-up and drop-off points, and the zone center GPS location is used instead in this project to replace zone IDs. Zone IDs are defined in a mysterious order and is not useful to represent geometric information, see Figure \ref{fig_manhattanZones} for more details.
\end{itemize}
\begin{figure}
    \centering
    \includegraphics[trim={0.2cm 0.7cm 2.4cm 2.9cm},clip,width=\linewidth]{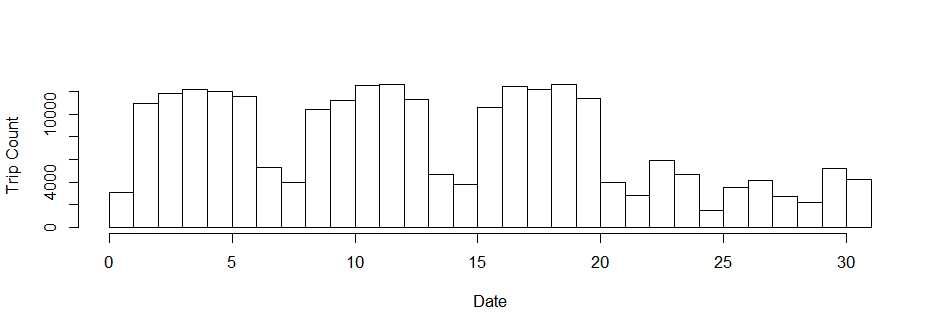}
    \caption{Histogram of Trip Dates (2019 Dec)}
    \label{fig_histDates}
\end{figure}
The data count is then reduced to 237727 or a $93\%$ pass rate. Using this cleaned data set, number of trips versus date is shown in Figure \ref{fig_histDates}. A weekly pattern is observed for the first three weeks (Dec 1 is Sunday). The number of trips are reduced significantly during the weekend. The pattern for the holiday season is quite different. In general, number of trips could be an indicator of traffic condition, and the trip time during the holiday season could be significantly away from the weekly pattern found in the first three weeks. As a result, all trips after December 22 are discarded and factor variable indicating day of week is added to the data set. This brings the total data count down to 200761. 

\subsection{Passenger Interarrival Time Distribution }
Another important part of the data analysis is to check passenger interarrival distribution. The key observation from the data set is that higher demand usually implies a better exponential distribution fit for the data. Data from the workdays of the first three weeks in December 2019 is used for the analysis. Demands for each zone is calculated by the total number of pick-ups from the corrected data set, and are listed below in Table \ref{tab:demand} in decreasing order.

\begin{table}[h]
\centering
\scalebox{0.56}{
\begin{tabular}{l l l l l l l l l l l l l l l l l l}
\hline
ID &236 &237 &186 &170 &141 &162 &140 &238 &142 &229 &239 &48 &161 &107 &263 &262 &234 \\
Demand &13243 &11026  &7296  &7182  &6762  &6534  &6350  &6090  &5669  &5576  &5426  &5221  &5173  &5126  &4775  &4467  &4166\\
ID &68 &100 &143 &230 &43 &233 &79 &113 &231 &164 &163 &249 &137    &90 &13 &246 &151\\ 
Demand  &3960 &3856 &3737 &3657 &3388 &3367 &3357 &3268 &3250 &3224 &3067 &3044 &2946 &2891 &2800 &2659 &2565\\ 
ID &75    &50   &158   &166    &87   &114    &41    &74   &125    &24   &148   &144   &211   &261    &42    &88   &116\\ 
Demand &2322  &2047  &2020  &1405  &1336  &1236  &1231  &1128  &1058   &974  &848   &794   &749   &575   &501   &458   &414\\
ID  &224     &4   &244   &209   &152   &232    &45\\ 
Demand &411   &371   &309   &289   &273   &223   &155 \\
\hline
\end{tabular}
}
\caption{Zone ID and the number of passengers taking a taxi between 8-9 AM on weekdays for three weeks of December 2019.}
\label{tab:demand}
\end{table}

\begin{figure}
    \centering
    \includegraphics[width=\textwidth]{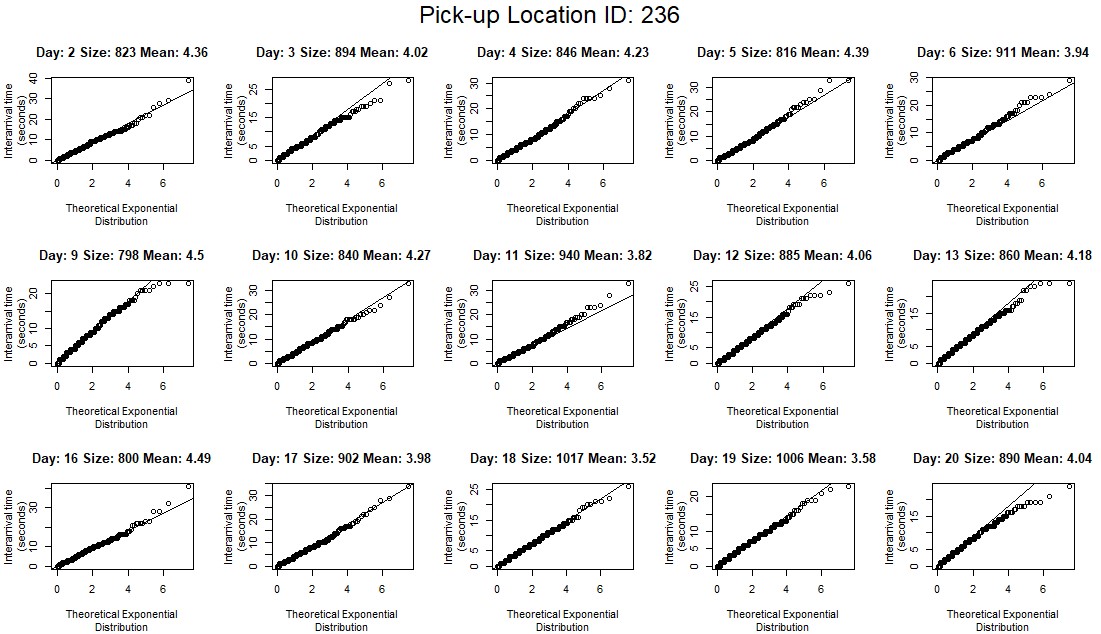}
    \centering
    \caption{Quantile-Quantile Plot for Zone ID 236 (High demand).}
    \label{fig:pickup_location_236}
\end{figure}

\begin{figure}
    \centering
    \includegraphics[width=\textwidth]{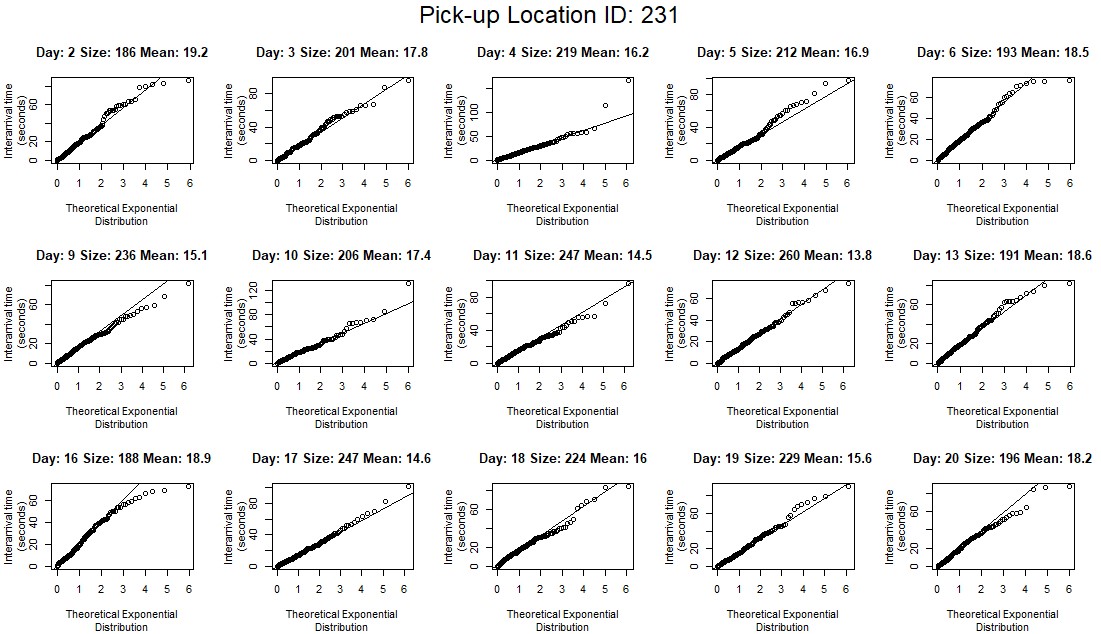}
    \centering
    \caption{Quantile-Quantile Plot for Zone ID 231 (Medium demand).}
    \label{fig:pickup_location_231}
\end{figure}

\begin{figure}
    \centering
    \includegraphics[width=\textwidth]{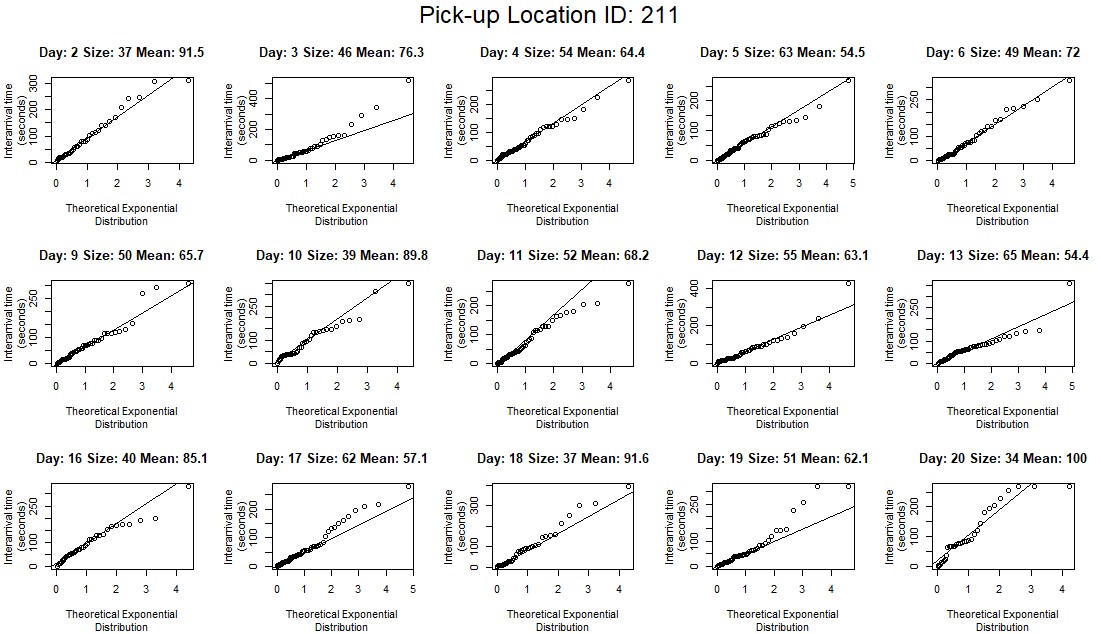}
    \centering
    \caption{Quantile-Quantile Plot for Zone ID 211 (Low demand).}
    \label{fig:pickup_location_211}
\end{figure}

The fitness of exponential distribution is then checked using quantile-quantile plot (qq-plot). To see the relationship of fitness and demand, location ID 236, 231, 211 are chosen to represent high, medium, and low demand. The qq-plots for these three locations are shown in Figures \ref{fig:pickup_location_236}, \ref{fig:pickup_location_231} and \ref{fig:pickup_location_211}. From the plots, we see that the qq-plots with ID 236 (high demand) have a relatively better fit for exponential distribution except for a few points with large values, which are mostly present in the data collected on December 20th. The mean interarrival time also has a relatively small deviation across the weekdays of the first three weeks of December. For the medium demand case (ID 231), the qq-plot looks similar, but the mean interarrival time for each day has a larger deviation compared with the high demand case. For the low demand case (ID 211), we notice that both exponential distribution and the constant mean assumptions are violated. To simplify the simulation model, we assume that top 20 locations with highest demands follow an exponential distribution with a constant rate (a lookup matrix) under the considered scenario.

\subsection{Data for Simulation}
Based on the corrected data, we output two matrices to the simulator. The first matrix is the average interarrival matrix. The row of the matrix is the location for pick up and the column of the matrix is the location for dropoff. The element in the $i$th row and $j$th column of the matrix is the mean interarrival time of passengers who want to travel from location $i$ to $j$, in seconds. The second matrix is the average trip matrix which has the same dimension as the first one but the element in the $i$th row and $j$th column of the matrix is the mean trip time of passengers who want to have a trip from location $i$ to $j$, also in seconds.

{}

{}

{}
\end{document}